\documentclass[english,prl, twocolumn, superscriptaddress]{revtex4-2}
\usepackage[T1]{fontenc}
\usepackage[latin9]{inputenc}
\usepackage{graphicx}
\usepackage{esint}

\makeatletter
\usepackage[colorlinks,citecolor=blue,linkcolor=blue]{hyperref}

\makeatother

\usepackage{babel}
\begin{document}
\title{Experimental implementation of finite-time Carnot cycle}
\date{\today}
\author{Ruo-Xun Zhai}
\address{Graduate School of China Academy of Engineering Physics, No. 10 Xibeiwang
East Road, Haidian District, Beijing, 100193, China}
\author{Fang-Ming Cui}
\address{Graduate School of China Academy of Engineering Physics, No. 10 Xibeiwang
East Road, Haidian District, Beijing, 100193, China}
\address{Beijing Normal University, Beijing 100875, China}
\author{Yu-Han Ma}
\address{Graduate School of China Academy of Engineering Physics, No. 10 Xibeiwang
East Road, Haidian District, Beijing, 100193, China}
\author{C. P. Sun}
\email{cpsun@gscaep.ac.cn}

\address{Graduate School of China Academy of Engineering Physics, No. 10 Xibeiwang
East Road, Haidian District, Beijing, 100193, China}
\address{Beijing Computational Science Research Center, Beijing 100193, China}
\author{Hui Dong}
\email{hdong@gscaep.ac.cn}

\address{Graduate School of China Academy of Engineering Physics, No. 10 Xibeiwang
East Road, Haidian District, Beijing, 100193, China}
\begin{abstract}
The Carnot cycle is a prototype of ideal heat engine to draw mechanical
energy from the heat flux between two thermal baths with the maximum
efficiency, dubbed as the Carnot efficiency $\eta_{\mathrm{C}}$.
Such efficiency can only be reached by thermodynamical equilibrium
processes with infinite time, accompanied unavoidably with vanishing
power - energy output per unit time. In real-world applications, the
quest to acquire high power leads to an open question whether a fundamental
maximum efficiency exists for finite-time heat engines with given
power. We experimentally implement a finite-time Carnot cycle with
sealed dry air as working substance and verify the existence of a
tradeoff relation between power and efficiency. Efficiency up to $(0.524\pm0.034)\eta_{\mathrm{C}}$
is reached for the engine to generate the maximum power, consistent
with the theoretical prediction $\eta_{\mathrm{C}}/2$. Our results
shall provide a new platform for studying finite-time thermodynamics
consisting of nonequilibrium processes.
\end{abstract}
\maketitle
\narrowtext

Heat engine is a major machinery to convert heat into the useful energy,
such as mechanic work or electricity. Sadi Carnot derived in 1824
an upper bound $\eta_{\mathrm{C}}=1-T_{\mathrm{c}}/T_{\mathrm{h}}$
\citep{Carnot1890,HerbertCallen1985} for the conversion efficiency
from heat to useful energy in the Carnot cycle with two isothermal
processes with the working substance in two thermal baths with temperatures
$T_{\mathrm{c}}$ and $T_{\mathrm{h}}$ and two adiabatic processes.
In isothermal processes, a control parameter is tuned in a quasi-static
fashion - physically much slower than the equilibrium time scale -
to reach the Carnot efficiency \citep{HerbertCallen1985}. Faster
processes typically result in more energy dissipation with a consequently
lower efficiency, yet can potentially increase the output power \citep{andresen1984thermodynamics}.
Such tradeoff between power and efficiency suggests the possibility
to thermodynamically optimize the two, e.g. increasing efficiency
while retaining the power or vise versa \citep{Novikov_1958,CA,Andresen1977,BroeckPRL2005,SeifertEPLStochatic,EspositoPRL2010,Tu2008JPhysAMathTheor41_312003}.
One might wonder what is the best achievable efficiency, if exists,
for any output power posted by the fundamental laws of thermodynamics
\citep{linearModelHulobec,TradeoffrelationShiraishi,Constraintrelationyhma}.

Answering such question requires the quantitative evaluation of irreversibility
in the fundamental non-equilibrium thermodynamics \citep{Prigogine1968book,StochasticThermodynamics}.
Theoretical models near equilibrium were explored recently to reveal
a tradeoff relation between power and efficiency \citep{Yan1989JCP,TradeoffrelationShiraishi,linearModelHulobec,Constraintrelationyhma,chenquan2022arxiv}.
More importantly, a fundamental limit of efficiency with a leading
order $\eta_{\mathrm{C}}/2$ is predicted for the engine generating
the maximum power \citep{BroeckPRL2005,Tu2008JPhysAMathTheor41_312003,EspositoPRL2010,tradeoffholubec}.
Aside from the theoretical achievements, it remains with urgency to
devise finite-time heat engine cycles to experimental{\large{}ly}
test the fundamental finite-time thermodynamical constraints \citep{MicrosizeHENatPhysics2011,BrownianHENatPhys2015,Rossnagel2016}.
The current main difficulty to implement a finite-time Carnot cycle
is to promptly change the temperature of the thermal bath before and
after the adiabatic processes, whose duration is far shorter than
the equilibrium time to avoid heat exchange \citep{BrownianHENatPhys2015}.
We develop an adiabatic-without-run scheme to implement the finite-time
Carnot cycle by separately running only two isothermal processes in
the high- and low-temperature baths. Without actual run of the two
adiabatic processes, we can change the bath temperature with the desired
accuracy to evaluate the performance of the finite-time Carnot engine.
Our scheme allows all the essential quantities for evaluating the
engine's performance to be obtained by the directly measurable work
in the two isothermal processes with the first law of thermodynamics
of energy conservation.

We present an experimental verification of the power-efficiency tradeoff
relation with the dry air as working substance in the finite-time
Carnot cycle, which is implemented by changing the gas volume $V(t)=V_{0}+\mathcal{A}L(t)$
via moving a piston along a designed path $L(t)$. $\mathcal{A}=\pi d^{2}/4$
is the area of the cross section of the cylindric chamber with the
diameter $d=5.00\mathrm{cm}$. We trace the pressure $p(t)$ with
pressure sensors on the chamber. The maintenance of the bath temperature
($\pm0.1\mathrm{K}$) is achieved by a designed water tank equipped
with the feedback temperature control unit.

\begin{figure*}
\includegraphics{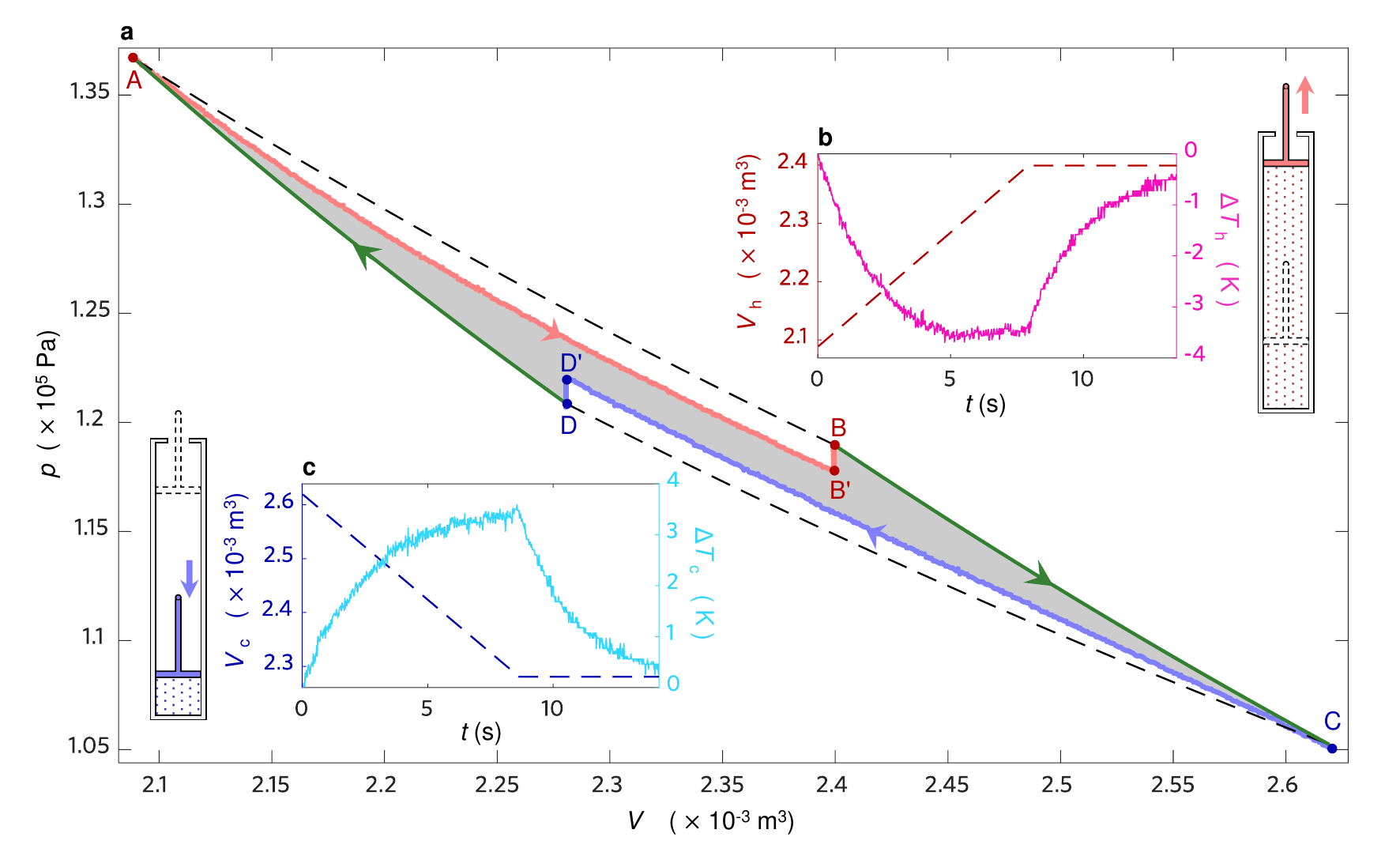}

\caption{The finite-time Carnot cycle under the temperature combination ($T_{\mathrm{h}},T_{\mathrm{c}}$)
$=(319.2.0,308.2)\mathrm{K}$. a. The Clapeyron pressure-volume ($p-V$)
graph of the finite-time Carnot cycle. The red and blue lines show
the finite-time isothermal expansion ($A\rightarrow B'\rightarrow B$)
with duration $\tau_{\mathrm{h}}$ and compression ($C\rightarrow C'\rightarrow D$)
with duration $\tau_{\mathrm{c}}$, and the green lines show the adiabatic
processes ($B\rightarrow C$ and $D\rightarrow A$) without actual
run. Two relaxation processes ($B'\rightarrow B$ and $D'\rightarrow D$)
with the waiting time $2\tau_{\mathrm{relax}}$ are added to allow
the system to reach the equilibrium with the thermal bath. The ideal
Carnot cycle ($A\rightarrow B\rightarrow C\rightarrow D$) is presented
as the gray dashed lines for reference. b-c. The gaseous volume $V_{\mathrm{h(c)}}(t)$
and the effective temperature trace $\Delta T_{\mathrm{h(c)}}(t)$
during the finite-time isothermal expansion and compression. The ideal
gas is expanded or compressed with the constant speed $L(t)=L_{0}+\nu t$,
with $\nu_{\mathrm{exp}}=20.00\mathrm{mm/s}$ and $\nu_{\mathrm{com}}=20.00\mathrm{mm/s}$
. The total time for the example cycle is $t_{\mathrm{tot}}=25.92\mathrm{s}$.}

\label{fig:setup}
\end{figure*}

The Clapeyron pressure-volume ($p-V$) graph is shown in Fig. \ref{fig:setup}a
with two finite-time isothermal processes (red and blue curves) and
two no-run adiabatic processes (green curves). In each run, the system
is immersed in the water bath for $10$ seconds to allow the initial
equilibration of the gas system with the water bath. And the gas is
expanded (red line $A\rightarrow B'$ in Fig. \ref{fig:setup}b) or
compressed (blue line $C\rightarrow D'$ in Fig. \ref{fig:setup}c)
in the two isothermal processes with constant speeds controlled by
a step motor with the precision $\pm0.02\mathrm{mm}$. The pressure
traces $p(t)$ (red and blue curves) measured in the two processes
deviates significantly from the equilibrium pressure (dashed black
curve) due to the finite relaxation time $\tau_{\mathrm{relax}}=2.77\mathrm{s}$.
After the expansion and compression, additional waiting time ($2\tau_{\mathrm{relax}}$)
is added to allow the gas relaxing to the equilibrium state ($B'\rightarrow B$
and $D'\rightarrow D$). The four piston positions ($A,B,C,D$) are
designed (see supplementary materials) for each temperature combination
($T_{\mathrm{h}}$,$T_{\mathrm{c}}$) to ensure the connection between
the end ($B$ and $D$) of one finite-time isothermal process to the
beginning ($C$ and $A$) of the other one with adiabatic processes.
We run the engine cycles with 10 temperature combinations to span
the Carnot efficiency from $\eta_{\mathrm{C}}=$0.0063 with the temperature
combination $(313.2,311.2)\mathrm{K}$ to $\eta_{\mathrm{C}}=0.0343$
with $(319.2.0,308.2)\mathrm{K}$.

\begin{figure*}
\includegraphics{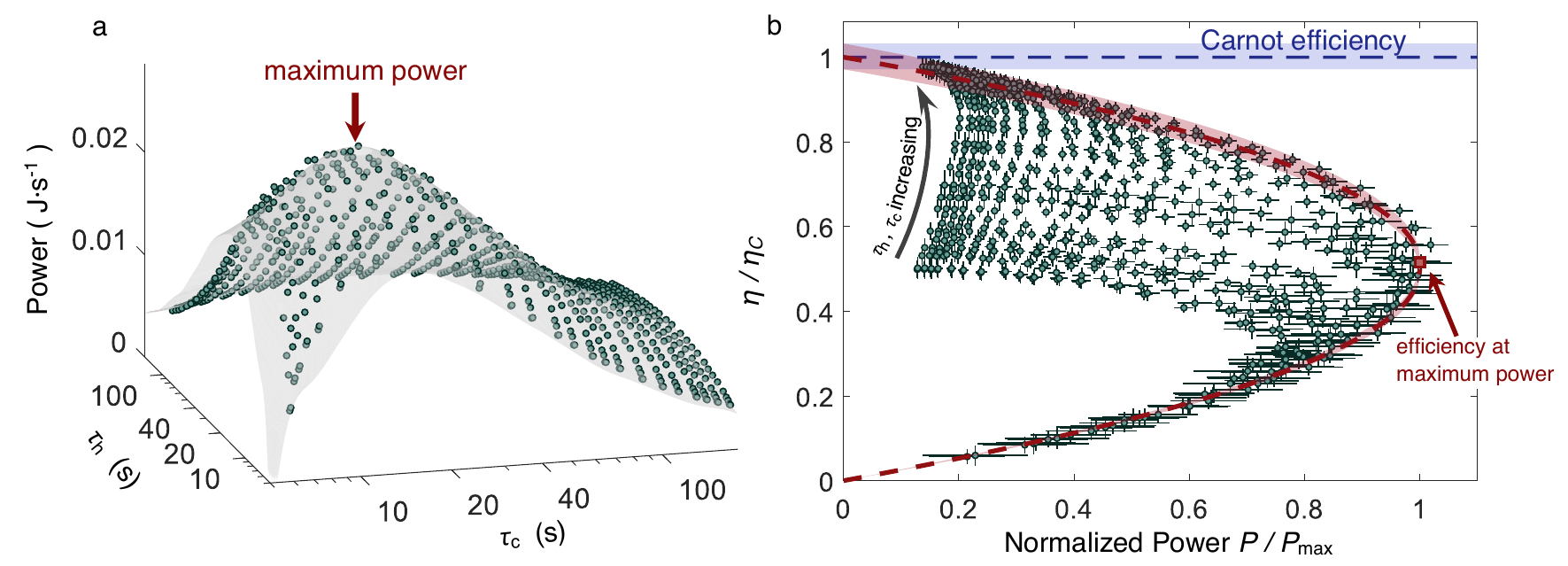}

\caption{Power and efficiency of the finite-time Carnot engine under the temperature
combination ($T_{\mathrm{h}},T_{\mathrm{c}}$) $=(319.2.0,308.2)\mathrm{K}$.
a. Output power $P(\tau_{\mathrm{h}},\tau_{\mathrm{c}})$ as a function
of operation time ($\tau_{\mathrm{h}},\tau_{\mathrm{c}}$). Each blue
circle shows the output power averaged over 8 repeats of the measurement.
The arrow shows the position of the maximum power with the control
time $\tau_{\mathrm{h}}^{*}=13.32\mathrm{s}$ and $\tau_{\mathrm{c}}^{*}=12.60\mathrm{s}$.
b. Efficiency-power tradeoff (blue circles with error bar). The dashed
green curve shows the Carnot efficiency $\eta_{\mathrm{C}}=0.0343$,
and the blue shadow presents the efficiency fluctuation due to temperature
variation of the water bath during the experiments. The dashed red
line shows the theoretical constraint \citep{Constraintrelationyhma}
between the normalized efficiency $\widetilde{\eta}\equiv\eta/\eta_{\mathrm{C}}$
and power $\widetilde{P}\equiv P/P_{\mathrm{max}}$ with the upper
bound $\widetilde{\eta}\protect\leq1-\left(1-\eta_{\mathrm{C}}\right)\widetilde{P}/[2\left(1+\sqrt{1-\widetilde{P}}\right)-\eta_{\mathrm{C}}\widetilde{P}]$
and the lower bound $\widetilde{\eta}\protect\geq1-\sqrt{1-\widetilde{P}}$
with the temperature fluctuation represented by the red shadow.}

\label{fig:PowerEff}
\end{figure*}

To evaluate the performance ( power and efficiency) of the designed
cycle, we calculate the work performed in each process with the integral
$W(t)=-\int p(t)\mathcal{A}dL(t)$. The heat exchange is obtained
with the conservation of energy as $\Delta Q=-W+\Delta U$, where
$\Delta U$ is the internal energy change of the gas. The important
properties of the ideal gas is that its internal energy depends only
on its temperature $T_{s}$, which is experimentally determined by
the ideal gas law via the state equation $pV=nRT_{\mathrm{s}}$ with
the amount of substance of gas in moles $n$ and the ideal gas constant
$R$. The temperature deviations from the thermal bath $\Delta T_{\mathrm{h(c)}}(t)=T_{\mathrm{s}}(t)-T_{\mathrm{h(c)}}$
are illustrated with solid lines (cyan and purple) in Fig. \ref{fig:setup}b
and \ref{fig:setup}c. The relaxation processes at the end of each
isothermal process ensure the unchanged internal energy $\Delta U_{A\rightarrow B}=\Delta U_{C\rightarrow D}=0$,
since the gas is nearly in equilibrium with the water bath $\Delta T_{\mathrm{h(c)}}(t)\approx0$.
The heat absorbed from the high(low)-temperature bath is measured
directly via 
\begin{equation}
Q_{\mathrm{h(c)}}=-W_{\mathrm{h(c)}}=\int_{0}^{\tau_{\mathrm{h(c)}}}p(t)\mathcal{A}\dot{L}(t)dt,
\end{equation}
where $\tau_{\mathrm{h}}$ ($\tau_{\mathrm{c}}$) is the operation
time for the isothermal expansion (compression) process $A\rightarrow B'\rightarrow B$
($C\rightarrow D'\rightarrow D$). The total work extracted $W_{\mathrm{tot}}$
of the whole cycle is measured by the difference of heat exchanges,
$W_{\mathrm{tot}}=Q_{\mathrm{h}}+Q_{\mathrm{c}}$. And the power is
calculated as the total work extracted during a cycle divided by the
total duration of the cycle, $P(\tau_{\mathrm{h}},\tau_{\mathrm{c}})=W_{\mathrm{tot}}$/($\tau_{\mathrm{h}}+\tau_{\mathrm{c}}$).
The efficiency is given by the ratio between the extracted work $W_{\mathrm{tot}}$
and the input heat $Q_{\mathrm{h}}$ from the high-temperature thermal
bath, $\eta(\tau_{\mathrm{h}},\tau_{\mathrm{c}})=W_{\mathrm{tot}}/Q_{\mathrm{h}}$.
In Fig. \ref{fig:PowerEff}a, we show how the power $P(\tau_{\mathrm{h}},\tau_{\mathrm{c}})$
changes with the two operation time $\tau_{\mathrm{h}}$ and $\tau_{\mathrm{c}}$
for one cycle under the temperature combination ($T_{\mathrm{h}},T_{\mathrm{c}}$)
$=(319.2.0,308.2)\mathrm{K}$. The competition between the increase
of work (equals to the gray area enclosed by the $p-V$ curve) and
the increase of operation time results in the maximum power $P_{\mathrm{max}}=0.030\mathrm{J}\cdot\mathrm{s}^{-1}$
on the power surface with the control time $\tau_{\mathrm{h}}^{*}=13.32\mathrm{s}$
and $\tau_{\mathrm{c}}^{*}=12.60\mathrm{s}$ (red arrow in Fig. \ref{fig:PowerEff}a).

Recently, much attentions are drawn to find the tradeoff between power
and efficiency for finite-time thermodynamic cycles, e.g., Carnot
cycle \citep{BroeckPRL2005,BrownianHENatPhys2015}. Within the framework
of the low-dissipation model \citep{BroeckPRL2005,EspositoPRL2010},
a power and efficiency tradeoff relation \citep{TradeoffrelationShiraishi,linearModelHulobec,Constraintrelationyhma}
is predicted for finite-time Carnot cycle. Figure \ref{fig:PowerEff}b
shows the scatter plot of the normalized efficiency $\eta/\eta_{\mathrm{C}}$
and the normalized output power $P/P_{\mathrm{max}}$ for cycles with
changing operation time $\tau_{\mathrm{h}}$ and $\tau_{\mathrm{c}}$.
The error bars of each set $P$ and $\eta$ are obtained from 8 repeats
of the experimental runs. The experimental data fall into the region
enclosed by two margins of maximum and minimum efficiencies, which
are illustrated as red dashed line with a red shadow area reflecting
the fluctuation of the bath temperature in Fig. \ref{fig:PowerEff}b.
The data illustrates not only an upper bound for the achievable efficiency,
but also a lower bound for the worst efficiency for the current finite-time
Carnot cycle. The cycle achieves the Carnot efficiency $\eta_{\mathrm{C}}=0.0343$
with the increasing operation time $\tau_{\mathrm{c}}$ and $\tau_{\mathrm{h}}$
at the top left corner with a vanishing power.

\begin{figure}
\includegraphics{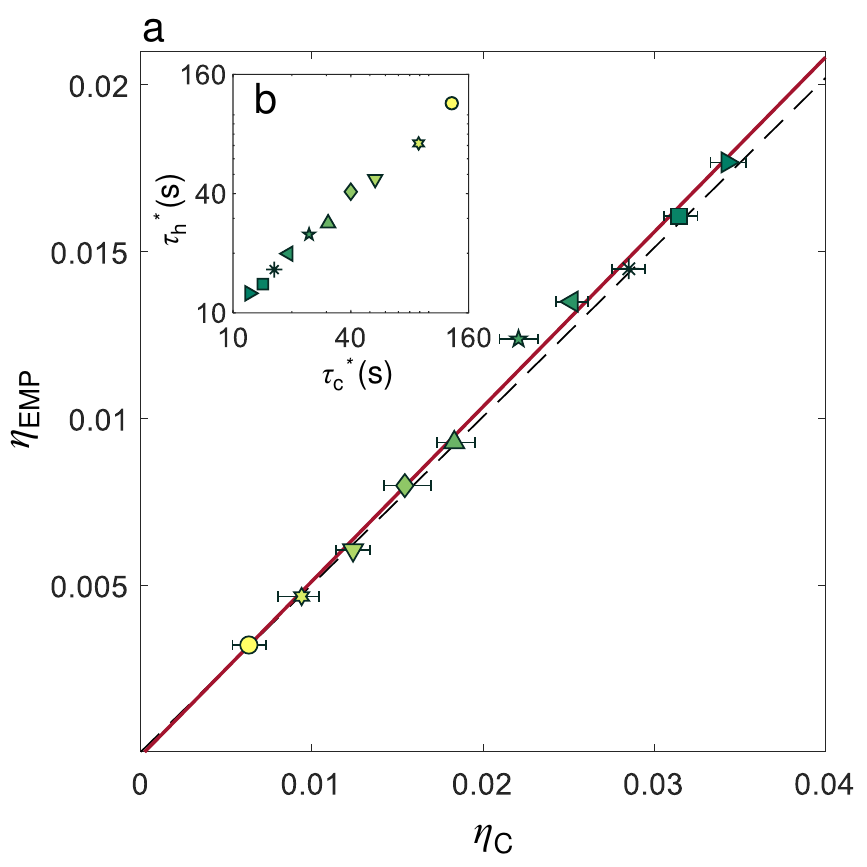}

\caption{Optimized finite-time Carnot cycle with maximum power. a. Efficiency
at maximum power (markers with error bars) as a function of Carnot
efficiency. The red solid line shows the linear fit $\eta_{\mathrm{EMP}}=a+b\eta_{\mathrm{C}}$
yielding the parameters $a=\pm7.0\times10^{-4}$ and $b=0.524\pm0.034$.
The dashed black line shows the Curzon-Ahlborn efficiency $\eta_{\mathrm{CA}}=1-\sqrt{1-\eta_{\mathrm{C}}}$.\textbf{
}b. Optimal operation time ($\tau_{\mathrm{h}}^{*},$$\tau_{\mathrm{c}}^{*}$)
for the maximum power in the cycles with different Carnot efficiencies.}

\label{fig:scaling}
\end{figure}

The key quantity to evaluate the finite-time Carnot cycle is the efficiency
at the maximum power $\eta_{\mathrm{EMP}}$, which was suggested as
a physical efficiency limit independent of the properties of the working
substance \citep{CA,EspositoPRL2010}. We extract the efficiency at
the maximum power $\eta_{\mathrm{EMP}}$ for all the temperature combinations
in our experiment, and show its dependence on the Carnot efficiency
in Fig. \ref{fig:scaling}a. The obtained maximum efficiencies (markers
with error bars) follow a simple relation $\eta_{\mathrm{EMP}}=(0.524\pm0.034)\eta_{\mathrm{C}}+\mathcal{O}(\eta_{\mathrm{C}}^{2})$,
which agrees well with the Curzon-Ahlborn efficiency\citep{CA} $\eta_{\mathrm{CA}}=1-\sqrt{1-\eta_{\mathrm{C}}}$
and the recent proposed bound \citep{EspositoPRL2010,SeifertEPLStochatic}
$\eta_{\mathrm{C}}/(2-\eta_{\mathrm{C}})$ to the first order of the
Carnot efficiency $\eta_{\mathrm{C}}$ as $\eta_{\mathrm{EMP}}=\eta_{\mathrm{C}}/2+\mathcal{O}(\eta_{\mathrm{C}}^{2})$.
The coefficient $1/2$ was proved as a universal value independent
of the system-specific features in the linear response regime due
to the symmetry of the Onsager relations \citep{BroeckPRL2005,StochasticThermodynamics}.
Our experimental data provides the first demonstration of the leading
order with the coefficient $1/2$.

The optimization of the cycle for maximum power is achieved by choosing
the operation time $\tau_{h}$ and $\tau_{c}$. We determine the corresponding
optimal operation time ($\tau_{\mathrm{h}}^{*},\tau_{\mathrm{c}}^{*}$)
(markers in Fig. \ref{fig:scaling}b) to reach the maximum power for
cycles with all temperature combinations in the current experiment.
The optimal time ($\tau_{\mathrm{h}}^{*},\tau_{\mathrm{c}}^{*}$)
is verified to be in the regime where the $1/\tau$ scaling is valid
\citep{BrownianHENatPhys2015,Ma2020} (discussions in the supporting
materials). With higher Carnot efficiency, less optimal operation
time is needed to achieve the maximum efficiency and in turn more
irreversibility is generated.

We have experimentally implemented the finite-time Carnot cycle with
the dry air as working substance. For any given output power, we have
shown the existence of the highest efficiency achieved with the designed
operation time. Our results verify the existence of a universal relation
of the efficiency at maximum power $\eta_{\mathrm{EMP}}=\eta_{\mathrm{C}}/2+\mathcal{O}(\eta_{\mathrm{C}}^{2})$
to the first order of the Carnot efficiency. We believe that our work
will spur more experimental efforts into explore the finite-time thermodynamics.

This work is supported by the National Natural Science Foundation
of China (NSFC) (Grants No. 12088101, No. 11534002, No. 11875049,
No. U1930402, No. U1930403 and No. 12047549) and the National Basic
Research Program of China (Grant No. 2016YFA0301201).

\bibliographystyle{naturemag}
\bibliography{EffPowExp}

\begin{thebibliography}{10}
\expandafter\ifx\csname url\endcsname\relax
  \def\url#1{\texttt{#1}}\fi
\expandafter\ifx\csname urlprefix\endcsname\relax\def\urlprefix{URL }\fi
\providecommand{\bibinfo}[2]{#2}
\providecommand{\eprint}[2][]{\url{#2}}

\bibitem{Carnot1890}
\bibinfo{author}{Carnot, S.}
\newblock \emph{\bibinfo{title}{Reflections on the Motive Power of Heat and on
  Machines Fitted to Develop that Power}} (\bibinfo{publisher}{J. Wiley},
  \bibinfo{year}{1890}).

\bibitem{HerbertCallen1985}
\bibinfo{author}{Callen, H.~B.}
\newblock \emph{\bibinfo{title}{Thermodynamics and an Introduction to
  Thermostatistics}} (\bibinfo{publisher}{John Wiley \& Sons},
  \bibinfo{year}{1985}).

\bibitem{andresen1984thermodynamics}
\bibinfo{author}{Andresen, B.}, \bibinfo{author}{Salamon, P.} \&
  \bibinfo{author}{Berry, R.~S.}
\newblock \bibinfo{title}{Thermodynamics in finite time}.
\newblock \emph{\bibinfo{journal}{Physics today}} \bibinfo{pages}{63}
  (\bibinfo{year}{1984}).

\bibitem{Novikov_1958}
\bibinfo{author}{Novikov, I.}
\newblock \bibinfo{title}{The efficiency of atomic power stations}.
\newblock \emph{\bibinfo{journal}{J. Nucl. Energy}}
  \textbf{\bibinfo{volume}{7}}, \bibinfo{pages}{125--128}
  (\bibinfo{year}{1958}).

\bibitem{CA}
\bibinfo{author}{Curzon, F.~L.} \& \bibinfo{author}{Ahlborn, B.}
\newblock \bibinfo{title}{Efficiency of a carnot engine at maximum power
  output}.
\newblock \emph{\bibinfo{journal}{Am. J. Phys.}} \textbf{\bibinfo{volume}{43}},
  \bibinfo{pages}{22--24} (\bibinfo{year}{1975}).

\bibitem{Andresen1977}
\bibinfo{author}{Andresen, B.}, \bibinfo{author}{Berry, R.~S.},
  \bibinfo{author}{Nitzan, A.} \& \bibinfo{author}{Salamon, P.}
\newblock \bibinfo{title}{Thermodynamics in finite time. i. the step-carnot
  cycle}.
\newblock \emph{\bibinfo{journal}{Phys. Rev. A}} \textbf{\bibinfo{volume}{15}},
  \bibinfo{pages}{2086--2093} (\bibinfo{year}{1977}).

\bibitem{BroeckPRL2005}
\bibinfo{author}{{Van den Broeck}, C.}
\newblock \bibinfo{title}{Introduction to thermodynamics of irreversible
  processes}.
\newblock \emph{\bibinfo{journal}{Phys. Rev. Lett.}}
  \textbf{\bibinfo{volume}{95}}, \bibinfo{pages}{190602}
  (\bibinfo{year}{2005}).

\bibitem{SeifertEPLStochatic}
\bibinfo{author}{Schmiedl, T.} \& \bibinfo{author}{Seifert, U.}
\newblock \bibinfo{title}{Efficiency at maximum power: An analytically solvable
  model for stochastic heat engines}.
\newblock \emph{\bibinfo{journal}{{EPL} (Europhysics Letters)}}
  \textbf{\bibinfo{volume}{81}}, \bibinfo{pages}{20003} (\bibinfo{year}{2007}).

\bibitem{EspositoPRL2010}
\bibinfo{author}{Esposito, M.}, \bibinfo{author}{Kawai, R.},
  \bibinfo{author}{Lindenberg, K.} \& \bibinfo{author}{{Van den Broeck}, C.}
\newblock \bibinfo{title}{Efficiency at maximum power of low-dissipation carnot
  engines}.
\newblock \emph{\bibinfo{journal}{Phys. Rev. Lett.}}
  \textbf{\bibinfo{volume}{105}}, \bibinfo{pages}{150603}
  (\bibinfo{year}{2010}).

\bibitem{Tu2008JPhysAMathTheor41_312003}
\bibinfo{author}{Tu, Z.~C.}
\newblock \bibinfo{title}{Efficiency at maximum power of
  feynman{\textquotesingle}s ratchet as a heat engine}.
\newblock \emph{\bibinfo{journal}{J. Phys. A: Math. Theor.}}
  \textbf{\bibinfo{volume}{41}}, \bibinfo{pages}{312003}
  (\bibinfo{year}{2008}).

\bibitem{linearModelHulobec}
\bibinfo{author}{Ryabov, A.} \& \bibinfo{author}{Holubec, V.}
\newblock \bibinfo{title}{Maximum efficiency of steady-state heat engines at
  arbitrary power}.
\newblock \emph{\bibinfo{journal}{Phys. Rev. E}} \textbf{\bibinfo{volume}{93}},
  \bibinfo{pages}{050101} (\bibinfo{year}{2016}).

\bibitem{TradeoffrelationShiraishi}
\bibinfo{author}{Shiraishi, N.}, \bibinfo{author}{Saito, K.} \&
  \bibinfo{author}{Tasaki, H.}
\newblock \bibinfo{title}{Universal trade-off relation between power and
  efficiency for heat engines}.
\newblock \emph{\bibinfo{journal}{Phys. Rev. Lett.}}
  \textbf{\bibinfo{volume}{117}}, \bibinfo{pages}{190601}
  (\bibinfo{year}{2016}).

\bibitem{Constraintrelationyhma}
\bibinfo{author}{Ma, Y.-H.}, \bibinfo{author}{Xu, D.}, \bibinfo{author}{Dong,
  H.} \& \bibinfo{author}{Sun, C.-P.}
\newblock \bibinfo{title}{Universal constraint for efficiency and power of a
  low-dissipation heat engine}.
\newblock \emph{\bibinfo{journal}{Phys. Rev. E}} \textbf{\bibinfo{volume}{98}},
  \bibinfo{pages}{042112} (\bibinfo{year}{2018}).

\bibitem{Prigogine1968book}
\bibinfo{author}{Prigogine, I.}
\newblock \emph{\bibinfo{title}{Introduction to the Thermodynamics of
  Irreversible Processes}} (\bibinfo{publisher}{Wiley}, \bibinfo{year}{1968}),
  \bibinfo{edition}{3rd edition} edn.

\bibitem{StochasticThermodynamics}
\bibinfo{author}{Seifert, U.}
\newblock \bibinfo{title}{Stochastic thermodynamics, fluctuation theorems and
  molecular machines}.
\newblock \emph{\bibinfo{journal}{Rep. Prog. Phys}}
  \textbf{\bibinfo{volume}{75}}, \bibinfo{pages}{126001}
  (\bibinfo{year}{2012}).

\bibitem{Yan1989JCP}
\bibinfo{author}{Chen, L.} \& \bibinfo{author}{Yan, Z.}
\newblock \bibinfo{title}{The effect of heat-transfer law on performance of a
  two-heat-source endoreversible cycle}.
\newblock \emph{\bibinfo{journal}{J. Chem. Phys.}}
  \textbf{\bibinfo{volume}{90}}, \bibinfo{pages}{3740--3743}
  (\bibinfo{year}{1989}).

\bibitem{chenquan2022arxiv}
\bibinfo{author}{Chen, Y.~H.}, \bibinfo{author}{Chen, J.-F.},
  \bibinfo{author}{Fei, Z.} \& \bibinfo{author}{Quan, H.~T.}
\newblock \bibinfo{title}{A microscopic theory of curzon-ahlborn heat engine}.
\newblock \emph{\bibinfo{journal}{arXiv}}  (\bibinfo{year}{2021}).

\bibitem{tradeoffholubec}
\bibinfo{author}{Holubec, V.} \& \bibinfo{author}{Ryabov, A.}
\newblock \bibinfo{title}{Maximum efficiency of low-dissipation heat engines at
  arbitrary power}.
\newblock \emph{\bibinfo{journal}{J. Stat. Mech.: Theory E.}}
  \textbf{\bibinfo{volume}{2016}}, \bibinfo{pages}{073204}
  (\bibinfo{year}{2016}).

\bibitem{MicrosizeHENatPhysics2011}
\bibinfo{author}{Blickle, V.} \& \bibinfo{author}{Bechinger, C.}
\newblock \bibinfo{title}{Realization of a micrometre-sized stochastic
  heat~engine}.
\newblock \emph{\bibinfo{journal}{Nat. Phys.}} \textbf{\bibinfo{volume}{8}},
  \bibinfo{pages}{143--146} (\bibinfo{year}{2011}).

\bibitem{BrownianHENatPhys2015}
\bibinfo{author}{Mart{\'{\i}}nez, I.~A.} \emph{et~al.}
\newblock \bibinfo{title}{Brownian carnot engine}.
\newblock \emph{\bibinfo{journal}{Nat. Phys.}} \textbf{\bibinfo{volume}{12}},
  \bibinfo{pages}{67--70} (\bibinfo{year}{2015}).

\bibitem{Rossnagel2016}
\bibinfo{author}{Rossnagel, J.} \emph{et~al.}
\newblock \bibinfo{title}{A single-atom heat engine}.
\newblock \emph{\bibinfo{journal}{Science}} \textbf{\bibinfo{volume}{352}},
  \bibinfo{pages}{325--329} (\bibinfo{year}{2016}).

\bibitem{Ma2020}
\bibinfo{author}{Ma, Y.-H.}, \bibinfo{author}{Zhai, R.-X.},
  \bibinfo{author}{Chen, J.}, \bibinfo{author}{Sun, C.~P.} \&
  \bibinfo{author}{Dong, H.}
\newblock \bibinfo{title}{Experimental test of the 1/t-scaling entropy
  generation in finite-time thermodynamics}.
\newblock \emph{\bibinfo{journal}{Phys. Rev. Lett.}}
  \textbf{\bibinfo{volume}{125}}, \bibinfo{pages}{210601}
  (\bibinfo{year}{2020}).

\end{thebibliography}


\begin{thebibliography}{1}
\expandafter\ifx\csname url\endcsname\relax
  \def\url#1{\texttt{#1}}\fi
\expandafter\ifx\csname urlprefix\endcsname\relax\def\urlprefix{URL }\fi
\providecommand{\bibinfo}[2]{#2}
\providecommand{\eprint}[2][]{\url{#2}}

\bibitem{BroeckPRL2005}
\bibinfo{author}{{Van den Broeck}, C.}
\newblock \bibinfo{title}{Introduction to thermodynamics of irreversible
  processes}.
\newblock \emph{\bibinfo{journal}{Phys. Rev. Lett.}}
  \textbf{\bibinfo{volume}{95}}, \bibinfo{pages}{190602}
  (\bibinfo{year}{2005}).

\bibitem{EspositoPRL2010}
\bibinfo{author}{Esposito, M.}, \bibinfo{author}{Kawai, R.},
  \bibinfo{author}{Lindenberg, K.} \& \bibinfo{author}{{Van den Broeck}, C.}
\newblock \bibinfo{title}{Efficiency at maximum power of low-dissipation carnot
  engines}.
\newblock \emph{\bibinfo{journal}{Phys. Rev. Lett.}}
  \textbf{\bibinfo{volume}{105}}, \bibinfo{pages}{150603}
  (\bibinfo{year}{2010}).

\bibitem{Krause2004Determining}
\bibinfo{author}{Krause, D.~E.} \& \bibinfo{author}{Keeley, W.~J.}
\newblock \bibinfo{title}{Determining the heat capacity ratio of air from
  almost adiabatic compressions}.
\newblock \emph{\bibinfo{journal}{Phys. Teach.}} \textbf{\bibinfo{volume}{42}},
  \bibinfo{pages}{481--483} (\bibinfo{year}{2004}).

\end{thebibliography}

\end{document}

\title{Supplementary material for ``Experimental implementation of finite-time
Carnot cycle''}
\date{\today}
\author{Ruo-Xun Zhai}
\affiliation{Graduate School of China Academy of Engineering Physics, Beijing 100193,
China}
\author{Fang-Ming Cui}
\affiliation{Graduate School of China Academy of Engineering Physics, Beijing 100193,
China}
\affiliation{Beijing Normal University, Beijing 100875 China}
\author{Yu-Han Ma}
\affiliation{Graduate School of China Academy of Engineering Physics, Beijing 100193,
China}
\author{C. P. Sun}
\email{cpsun@gscaep.ac.cn}

\affiliation{Graduate School of China Academy of Engineering Physics, Beijing 100193,
China}
\affiliation{Beijing Computational Science Research Center, Beijing 100193, China}
\author{Hui Dong}
\email{hdong@gscaep.ac.cn}

\affiliation{Graduate School of China Academy of Engineering Physics, Beijing 100193,
China}

\maketitle
%\narrowtext

\section{Experimental Apparatus - Calibration of the cylindric volume}

\textbf{Experimental setup.} The sealed dry air is prepared as the
work substance for the finite-time Carnot cycle. The dry detergent
($\mathrm{CaCl_{2}}$) is placed in the chamber before sealed with
the silicon grease to avoid gas leakage. The chamber (homebuilt) is
set for 24 hours to allow the absorption of the moisture in the sealed
air. The water bath ($\sim0.1\mathrm{m}^{3}$) is circulated with
a water pump to ensure the uniform temperature distribution in the
water bath. Before each sets of finite-time Carnot cycles, the water
temperature is firstly raised to the desired degree ($\pm0.1\mathrm{K}$).
And the chamber is then immersed for 30 minutes in the bath to allow
the initial equilibrium with the water bath.

Three pressure sensors measure the absolute pressure with the range
($0$-$0.15\mathrm{MPa}$) and the precision $0.2\%$ F.S ($300\mathrm{Pa}$).
The position sensor measures the piston displacement with the range
($0\mathrm{mm}$- $300.00\mathrm{mm}$). The obtained data is used
to synchronize the position data from stepper motor. The temperature
sensor (PT100) with the precision $0.1\mathrm{K}$ is used to monitor
the temperature of water bath. The data acquisition is through collecting
the current (4-20mA) at the frequency of 50Hz.

The piston is controlled by the stepper motor with the maximum speed
$v_{\mathrm{max}}=150.00\mathrm{mm/s}$ and maximum acceleration $a_{\mathrm{max}}=400.00\mathrm{mm/s}^{2}$.
The total time $\tau=\Delta L/v+v/a_{\mathrm{max}}+2\tau_{\mathrm{relax}}$
is archived with the designed speed $v$ to move the piston by the
displacement $\Delta L$.

\begin{figure}[H]
\centering{}\includegraphics{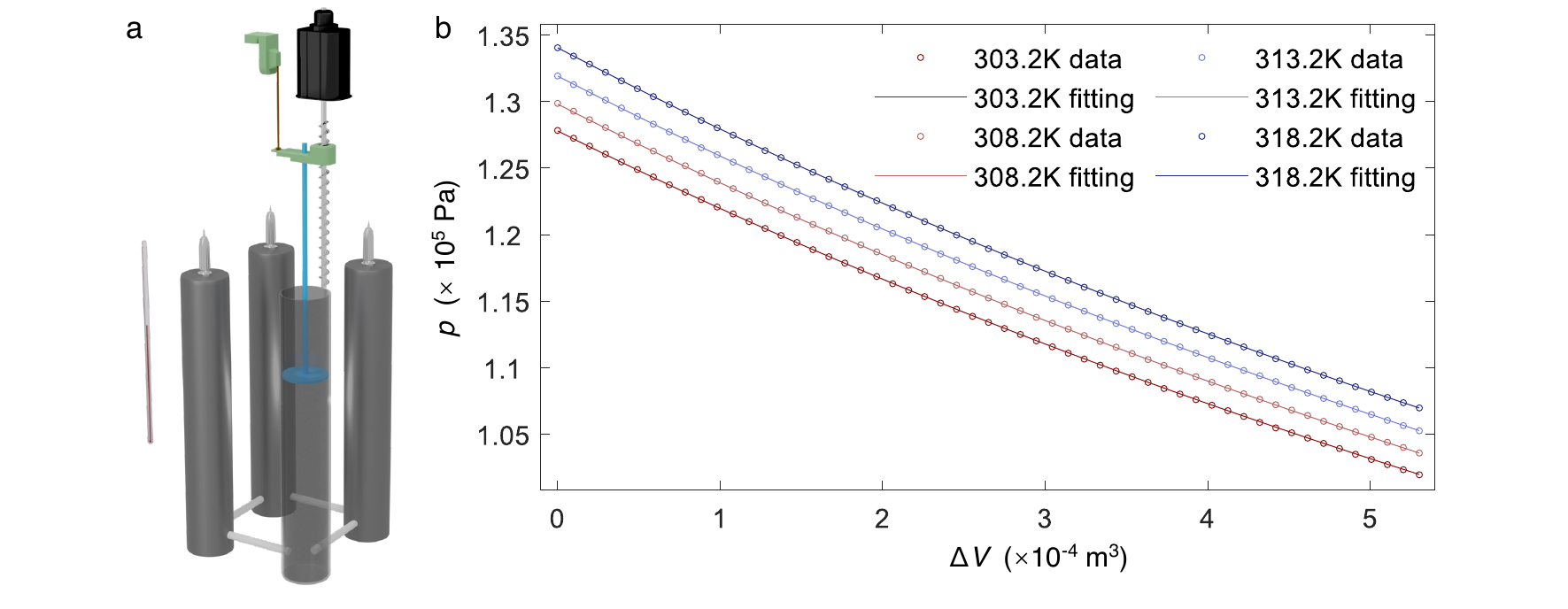} \caption{The experimental setup and the volume calibration. a. The experimental
setup. b. Fitting of ideal gas law to obtain the amount of substance
$n$ and the volume $V_{\mathrm{min}}$. We measured $\Delta V$ together
with $p$ when the bath is at $303.2\mathrm{K},308.2\mathrm{K},313.2\mathrm{K},$
and $318.2\mathrm{K}$. The fitting of the data with $p(V_{\mathrm{min}}+\Delta V)=nRT$
yields $V_{\mathrm{min}}=2.094\times10^{-3}\mathrm{m^{3}}$, $n=0.1062\mathrm{mol}$.
\label{fig:diagramFitting}}
\end{figure}

The design of the chamber ( diameter $d=5.00\mathrm{cm}$) is illustrated
in Figure~\ref{fig:diagramFitting}(a). With the area of the piston
section $\mathcal{A}=\pi d^{2}/4=1.963\times10^{-3}\mathrm{m^{3}}$,
we measure the gas volume change $\Delta V=\mathcal{A}\Delta L\equiv V-V_{\mathrm{min}}$
with $\Delta L$ as the displacement of stepper motor. The unknown
volume $V_{\mathrm{min}}$ is calibrated with the state function of
the ideal gas 
\begin{equation}
p(V_{\mathrm{min}}+\mathcal{A}\Delta L)=nRT,\label{eq:statefunction}
\end{equation}
where $p$, $\Delta L$ and $T$ are measured with sensors. Here $R=8.3145\mathrm{J\cdot K^{-1}}\cdot\mathrm{mol}^{-1}$
is the ideal gas constant, and $n$ is the amount of substance. For
one temperature, we measured the pressure $p$ as a function of the
displacement $\Delta L$. The piston is moved to one position and
set for $30$ seconds to allow the equilibrium in order to acquire
the pressure data $p(\Delta L)$. The data for the pressure $p$ as
a function of $\Delta V=\mathcal{A}\Delta L$ is illustrated in Fig.
\ref{fig:diagramFitting} for different temperatures $303.2\mathrm{K},308.2\mathrm{K},313.2\mathrm{K},$
and $318.2\mathrm{K}$. And the volume $V_{\mathrm{min}}$ and $n$
are treated as two fitting parameters. The fitting of Eq.\ref{eq:statefunction}
yields $V_{\mathrm{min}}=2.094\times10^{-3}\mathrm{m^{3}},n=0.1062\mathrm{mol}$.
The volume is then explicitly expressed as function of the displacement
of the piston $\Delta L$ as
\begin{equation}
V=1.963\times10^{-3}\mathrm{m^{3}}\times\Delta L+2.094\times10^{-3}\mathrm{m^{3}}.
\end{equation}

\section{Finite-time Carnot cycle design}

\begin{figure}
\includegraphics{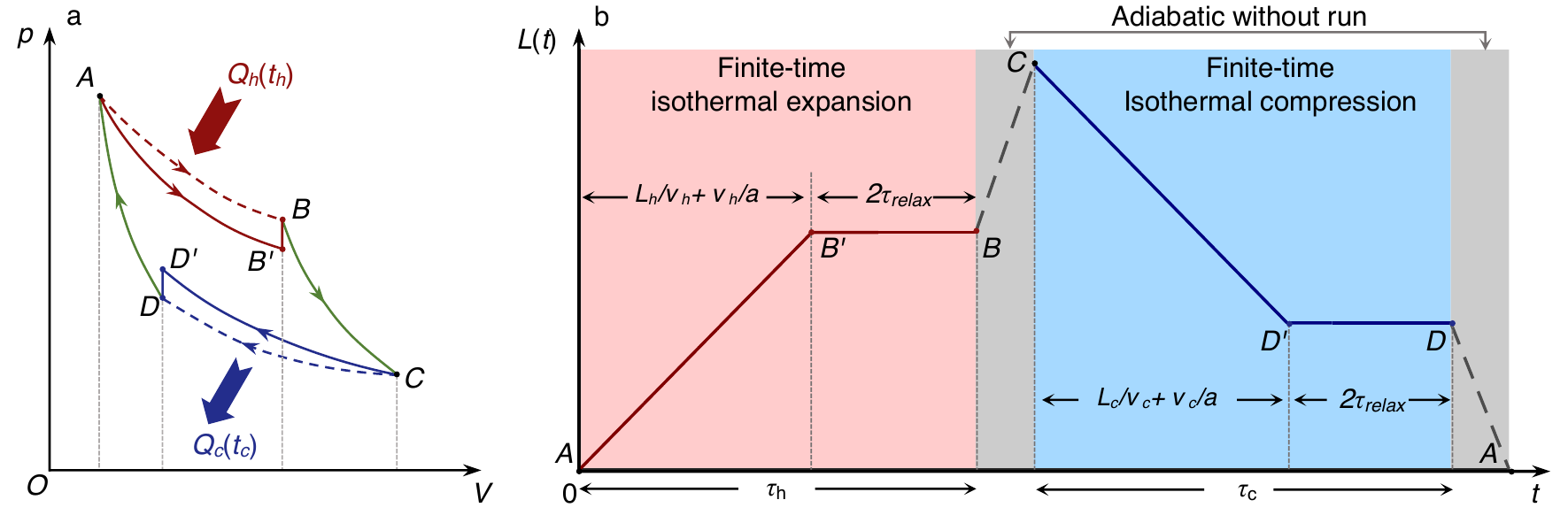}

\caption{Diagram of finite-time Carnot cycle in our adiabatic-without-run scheme
and the piston position $L(t)$. (a) The Clapeyron pressure-volume
($p(t)-V(t)$) diagram. (b) The designed piston movement $L(t)$ for
the cycle.}

\label{fig:AdiabaticWitoutRun}
\end{figure}

Our design of the adiabatic-without-run scheme is illustrated in Figure
\ref{fig:AdiabaticWitoutRun}. The key ideal is to complete the finite-time
Carnot cycle with only running two finite-time isothermal processes
and without running the two adiabatic processes. Such design is based
on the fact that the evaluation of the finite-time Carnot cycle is
through the two experimentally measured quantities, the total work
extracted and the total time of the cycle. The total work extracted
$W_{\mathrm{tot}}=Q_{\mathrm{h}}+Q_{\mathrm{c}}$ is obtained by the
difference of the heat exchange during the finite-time isothermal
expansion ($Q_{\mathrm{h}}$) and compression process ($Q_{\mathrm{c}}$)
with the help of the first law of thermodynamics - energy conservation.
Therefore there is no need to actually run the two adiabatic processes
to obtain the total work. In term of the total cycle time, the operation
of adiabatic process requires short operation time (shorter than the
relaxation time $\tau_{\mathrm{relax}}$) to avoid heat exchange.
The time for adiabatic processes are neglected in our experiments
as well as in the theoretical works \citep{BroeckPRL2005,EspositoPRL2010}. 

\textbf{Experimental protocol}. To complete the cycle in our adiabatic-without-run
scheme, we have to ensure the connection between the end of one finite-time
isothermal process to the beginning of the other one with an adiabatic
process via the following equations
\begin{align}
V_{B}^{\gamma-1}T_{\mathrm{h}} & =V_{C}^{\gamma-1}T_{\mathrm{c}},\nonumber \\
V_{D}^{\gamma-1}T_{\mathrm{c}} & =V_{A}^{\gamma-1}T_{\mathrm{h}},\label{eq:connection}
\end{align}
where $\gamma$ is the ratio between specific heats at constant pressure
and at constant volume ($\gamma=1.40$ for the dry air \citep{Krause2004Determining}).
The additional relaxation processes with the duration $2\tau_{\mathrm{relax}}$
are added to allow the equilibrium with the thermal bath. The minimum
of the volume $V_{\mathrm{min}}$ is reached with $L_{A}=0$, and
the maximum volume $V_{C}$ is reached at $L_{C}=270.00\mathrm{mm}$.
The sets of the designed connection points $L_{i}$ ($i=B,D$) are
calculated with the connection condition in Eq. (\ref{eq:connection})
and presented in Table \ref{tab:cycleparameters}. 

\begin{table}[H]
\begin{centering}
\begin{tabular}{ccccc>{\centering}p{0.5cm}cc}
\hline 
\multirow{2}{*}{$T_{\mathrm{h}}$$(\mathrm{K})$} & \multirow{2}{*}{$T_{\mathrm{c}}$$(\mathrm{K})$} & \multirow{2}{*}{$\eta_{C}$} & \multicolumn{2}{c}{Finite-time isothermal expansion} &  & \multicolumn{2}{c}{Finite-time isothermal compression}\tabularnewline
\cline{4-5} \cline{5-5} \cline{7-8} \cline{8-8} 
 &  &  & $L_{A}$(mm) & $L_{B}$(mm) &  & $L_{C}$(mm) & $L_{D}$(mm)\tabularnewline
313.2 & 311.2 & $6.39\times10^{-3}$ & 0 & 248.76 &  & 270.00 & 17.22\tabularnewline
313.2 & 310.2 & $9.58\times10^{-3}$ & 0 & 238.22 &  & 270.00 & 25.98\tabularnewline
314.2 & 310.2 & $12.73\times10^{-3}$ & 0 & 227.86 &  & 270.00 & 34.72\tabularnewline
315.2 & 310.2 & $15.87\times10^{-3}$ & 0 & 217.62 &  & 270.00 & 43.50\tabularnewline
316.2 & 310.2 & $18.98\times10^{-3}$ & 0 & 207.49 &  & 270.00 & 52.33\tabularnewline
317.2 & 310.2 & $22.07\times10^{-3}$ & 0 & 197.47 &  & 270.00 & 61.20\tabularnewline
317.2 & 309.2 & $25.22\times10^{-3}$ & 0 & 187.31 &  & 270.00 & 70.34\tabularnewline
318.2 & 309.2 & $28.29\times10^{-3}$ & 0 & 177.48 &  & 270.00 & 79.32\tabularnewline
319.2 & 309.2 & $31.33\times10^{-3}$ & 0 & 167.76 &  & 270.00 & 88.34\tabularnewline
319.2 & 308.2 & $34.47\times10^{-3}$ & 0 & 157.80 &  & 270.00 & 97.74\tabularnewline
\hline 
\end{tabular}
\par\end{centering}
\caption{Table for the piston positions of finite-time Carnot cycle. The piston
position $B$ and $D$ are calculated with Eq. (\ref{eq:connection})
.}

\label{tab:cycleparameters}
\end{table}

For each cycle, $27$ different values of speeds ($v_{\mathrm{h}}$
and $v_{\mathrm{c}}$) are chosen to achieve the series of control
time set $\tau_{\mathrm{h}}$ and $\tau_{\mathrm{c}}$ for finite-time
isothermal expansion and compression processes. The value of the these
speeds $v$ are 150.00, 120.00, 100.00, 80.00, 60.00, 40.00, 30.00,
24.00, 20.00, 15.00, 12.00, 10.00, 8.00, 6.00, 5.00, 4.00, 3.50, 3.00,
2.70, 2.40, 2.00, 1.70, 1.50, 1.30, 1.20, 1.10, and 1.00 $\mathrm{mm/s}$.
For each speed, the process are repeated for 8 times.

\begin{figure*}[!htbp]
\centering{}\includegraphics{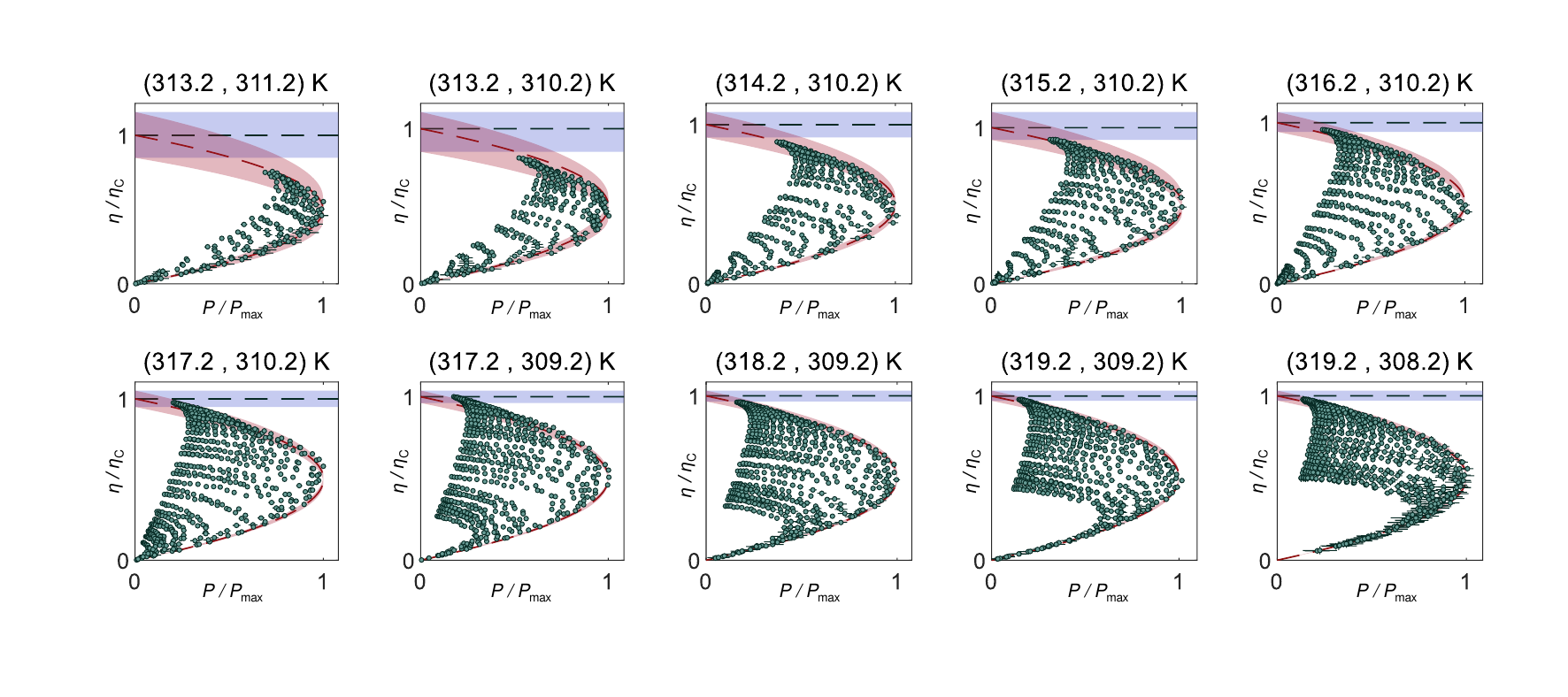} \caption{The $\eta-P$ constraint relation for all the temperature sets measured
by the sensor 1 and the strategy to find the efficiency at the maximum
power. Cycles with negative output power are not plotted in the above
figures. }
\label{fig:allsensors_constrain}
\end{figure*}

\section{Data Processing and Discussion}

The heat exchange with the water bath is obtained via calculating
the work in the finite-time isothermal expansion and compression processes,
i.e, 
\begin{align}
Q_{\mathrm{h}} & =-W_{A\rightarrow B'\rightarrow B}=-W_{\mathrm{h}}(\tau_{\mathrm{h}}),\\
Q_{\mathrm{c}} & =-W_{C\rightarrow D'\rightarrow D}=-W_{\mathrm{c}}(\tau_{\mathrm{c}}),
\end{align}
where $\tau_{\mathrm{h}}=(L_{B}-L_{A})/v_{\mathrm{h}}+v_{\mathrm{h}}/a_{\mathrm{max}}$
and $\tau_{\mathrm{c}}=(L_{C}-L_{D})/v_{\mathrm{c}}+v_{\mathrm{c}}/a_{\mathrm{max}}$.
Trapezoidal integrals is employed to compute the work performed in
a single process, 
\begin{equation}
W=\sum_{i}(V_{i+1}-V_{i})\frac{p_{i+1}+p_{i}}{2},
\end{equation}
where $V_{i}$ and $p_{i}$ are the volume and pressure data points.

The total work extracted is $W_{\mathrm{tot}}(\tau_{\mathrm{h}},\tau_{\mathrm{c}})=Q_{\mathrm{h}}+Q_{\mathrm{c}}$
by using the energy conservation law. The power and efficiency are
obtained as 
\begin{align}
P(\tau_{\mathrm{h}},\tau_{\mathrm{c}}) & =\frac{W_{\mathrm{tot}}(\tau_{\mathrm{h}},\tau_{\mathrm{c}})}{\tau_{\mathrm{h}}+\tau_{\mathrm{c}}},\\
\eta(\tau_{\mathrm{h}},\tau_{\mathrm{c}}) & =\frac{W_{\mathrm{tot}}(\tau_{\mathrm{h}},\tau_{\mathrm{c}})}{Q_{\mathrm{h}}}.
\end{align}

One set of the power-efficiency with temperature combination $\left(319.2,308.2\right)\mathrm{K}$
is shown in the main context. We present the scattering plots of the
power and the efficiency along with theoretical constraint relations
for all the temperature sets in Figure \ref{fig:allsensors_constrain}.
The current data are from the pressure sensor S1. There are $27\times27$
different combinations of $\tau_{\mathrm{h}}$ and $\tau_{\mathrm{c}}$.
The data with negative work are neglected in the current plot. The
data for all the temperatures fall into the theoretical predictions.

\begin{figure}
\includegraphics{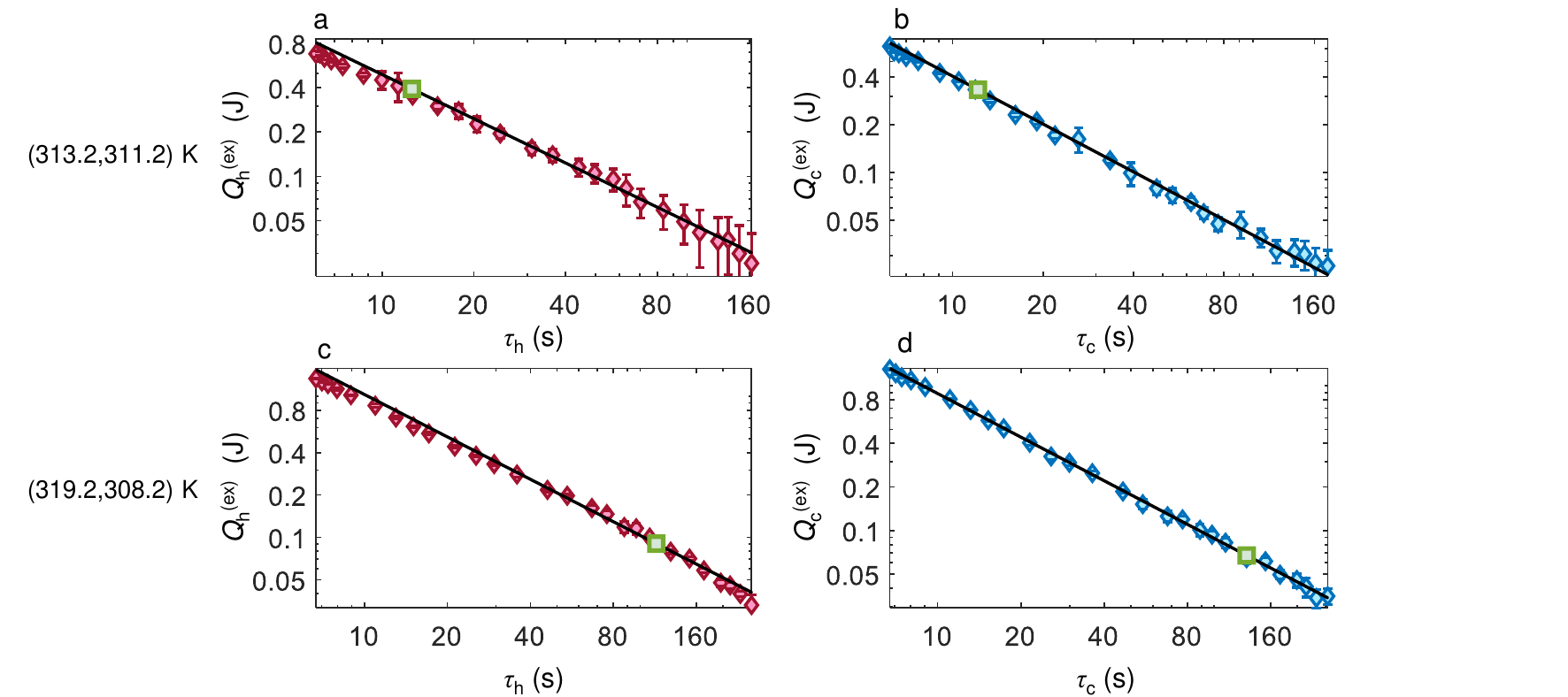}

\caption{The $1/\tau$ scaling of the low-dissipation model for the temperature
sets $\left(313.2,311.2\right)\mathrm{K}$ and $\left(319.2,308.2\right)\mathrm{K}$.
The optimal control time ($\tau_{\mathrm{h}}^{*}$ and $\tau_{\mathrm{c}}^{*}$)
is marked with green squares on the figures. }

\label{fig:optimaltime}
\end{figure}

We check the validity of the $1/\tau$ scaling of irreversibility
in the low-dissipation model. The irreversibility is measured as the
extra heat exchange $Q_{\mathrm{h(c)}}^{(\mathrm{ex})}=|Q_{\mathrm{h(c)}}(\tau_{\mathrm{h(c)}})-Q_{\mathrm{h(c)}}^{(\mathrm{quasi})}|$,
where $Q_{\mathrm{h(c)}}^{(\mathrm{quasi})}$ is the heat exchange
during the quasi-static isothermal processes ($A\rightarrow B$ and
$C\rightarrow D$). The extra heat exchange $Q_{\mathrm{h(c)}}^{(\mathrm{ex})}$
is plotted as functions of the operation time $\tau_{\mathrm{h(c)}}$
in Figure \ref{fig:optimaltime} for two temperature sets $\left(313.2,311.2\right)\mathrm{K}$
and $\left(319.2,308.2\right)\mathrm{K}$. We mark the optimal operation
time $\tau_{\mathrm{h}}^{*}$ and $\tau_{\mathrm{c}}^{*}$ for maximum
power on the plot with green squares. 

\bibliographystyle{naturemag}
\bibliography{EffPowExp}